\documentclass[aps,pre,preprint, showpacs, superscriptaddress]{revtex4}
\usepackage{latexsym}
\usepackage{natbib}
\usepackage{amsmath}
\usepackage{amsfonts}
\usepackage{amssymb}
\usepackage[utf8]{inputenc}
\usepackage[T1]{fontenc}
\usepackage[dvips]{color,graphicx}   
\usepackage{verbatim}   
\usepackage{t1enc}
\usepackage{anysize}
\usepackage{tocbibind}
\usepackage{subfig}
\usepackage[percent]{overpic} 
\usepackage{float}  
\usepackage{appendix} 
\usepackage{booktabs} 
\usepackage{natbib}
\usepackage{array}

\begin{document}
\title{Epidemic spreading on modular networks: The fear to declare a pandemic}
\author{Lucas D. Valdez}\email{}
\affiliation{Department of Physics, Boston University, Boston, Massachusetts 02215, USA}
\author{Lidia A. Braunstein}
\affiliation{Instituto de
  Investigaciones F\'isicas de Mar del Plata (IFIMAR)-Departamento de
  F\'isica, FCEyN, Universidad Nacional de Mar del Plata-CONICET, Mar
  del Plata 7600, Argentina.} 
\affiliation{Department of Physics, Boston University, Boston, Massachusetts 02215, USA}
\author{Shlomo Havlin}
\affiliation{Department of Physics, Bar Ilan University, Ramat Gan 5290002, Israel}
\affiliation{Department of Physics, Boston University, Boston, Massachusetts 02215, USA}
\affiliation{Tokyo Institute of Technology, Yokohama 152-8550, Japan}

\date{\today}

\begin{abstract}
In the past few decades, the frequency of pandemics has been increased due
to the growth of urbanization and mobility among countries. Since a
disease spreading in one country could become a pandemic with a
potential worldwide humanitarian and economic impact, it is important
to develop models to estimate the probability of a worldwide
pandemic. In this paper, we propose a model of disease spreading in a
structural modular complex network (having communities) and study how the
number of bridge nodes $n$ that connect communities affects
disease spread. We find that our model can be described at a global
scale as an infectious transmission process between communities with
global infectious and recovery time distributions that depend on the
internal structure of each community and $n$. We find that near the
critical point as $n$ increases, the disease reaches most of the
communities, but each community has only a small fraction of recovered
nodes. In addition, we obtain that in the limit $n \to \infty$, the
probability of a pandemic increases abruptly at the critical
point. This scenario could make the decision on whether to launch a
pandemic alert or not more difficult. Finally, we show that link
percolation theory can be used at a global scale to estimate the
probability of a pandemic since the global transmissibility between
communities has a weak dependence on the global recovery time.
\end{abstract}
\pacs{---}

\maketitle

\section{Introduction}
Community or modular structure is a ubiquitous property in real
complex networks that can be found in systems such as brain networks,
social networks, and technological
networks~\cite{fortunato2010community,alexander2012discovery,girvan2002community}. A
community is a sub-graph with more internal than external connections,
and as the number of internal links increases compared to the external
ones, the network has a higher level of community structure or
modularity~\cite{zhou2007phase,fortunato2010community}. Several
theoretical studies have focused on studying models of networks with
sub-graphs whose nodes are densely connected in order to understand
the effect of the community structure on processes that develop on top
of complex
networks~\cite{arenas2008synchronization,wang2019coevolution,nematzadeh2014optimal,lazaridis2018spontaneous}. Disease
spreading is one of the most studied dynamic processes since many
diseases that emerge could become an epidemic, i.e., could affect a
large number of people, or even could spread across the world and
become a pandemic. Nowadays, due to the enhanced human 
migration from rural to urban
regions~\cite{fields1975rural,zhang2003rural}, many people live in
agglomerated cities throughout the planet where the number of internal
contacts is much higher than the number of contacts among people from
different cities. When a disease spreads between different cities or
regions, it is essential for national and international health
authorities to activate mitigation or immunization strategies, when
a disease is a small outbreak, an epidemic, or even a
pandemic. Therefore, developing models is crucial to predict the
epidemic and pandemic potential of a disease spreading and for developing
mitigation strategies.

The susceptible-infected-recovered (SIR) model is widely used to study
diseases that confer permanent immunity~\cite{Anderson-1992}. In this
model, the nodes can be in one of the following states: 1)
susceptible, i.e., a node that is healthy but not immunized to the
disease, 2) infected, and thus can transmit the disease to its
susceptible neighbors, and 3) recovered, which is a node that cannot
transmit the disease because it acquired permanent immunity. For a
discrete-time evolution, the dynamic rules of the SIR model are as follows: An
infected individual tries to infect a susceptible neighbor with
probability $\beta$ per unit time step and recovers after a fixed
recovery time, $t_r$, that could be the same for all nodes or follow a
probability distribution
$P(t_r)$~\cite{valdez2012temporal,miller2018primer}. A relevant
parameter of this model is the transmissibility $T$, which is the
effective transmission or infection probability and depends on $\beta$
and $t_r$~\cite{newman2002spread}. At the initial state of the dynamic
process, all the nodes are susceptible except for one infected node
called the index case, from where the disease might spread throughout
the network. During the early stages of the dynamic process, there are
only a few infected nodes, and hence the process is in a stochastic
regime in which the disease could be halted due to fluctuations or
noise~\cite{allen2000comparison}. The disease reaches the final state
when it stops spreading, and there are only susceptible and/or
recovered nodes. In homogeneous networks with no community structure
in the thermodynamic limit, the disease becomes an epidemic if a
finite fraction of nodes is recovered, and it is an outbreak
otherwise. In the SIR model, there exists a critical value $T_c$ below
which the probability $\Pi$ of an epidemic is null, while for $T>T_c$,
$\Pi>0$. However, note that not necessarily $\Pi=1$ for $T>T_c$, so
the disease could end up in an outbreak due to the fluctuations in the
early dynamic, as mentioned above. Lagorio {\it et
  al.}~\cite{lagorio2009effects} showed that there exists a cutoff
$s_c$ of the size of the number of recovered nodes above which the
disease is in an epidemic state while below $s_c$ it is an
outbreak. Newman obtained that at the final state, the
transmissibility $T$ governs the fraction of recovered nodes which is
identical to the relative size of the giant component (GC) for a link
percolation process (with a probability of link occupation
$p=T$)~\cite{stauffer2014introduction,havlin1991fractals,newman2002spread}. In
turn, the SIR model exhibits a second-order transition at a critical
threshold $T_c$ which value coincides with the critical probability of
link occupation in a link percolation process. The outcome of a
disease does not depend only on the SIR parameters, $\beta$ and $t_r$,
but also on the network structure. Newman~\cite{newman2002spread}
showed that for a random homogeneous network (without communities) and
having degree distribution $P(k)$ (where $k$ is the connectivity or
the number of neighbors of a node), the critical transmissibility
$T_c$ depends on the first moment $\langle k \rangle$, and second
moment $\langle k^2 \rangle$ of the degree distribution. This is
analogous to the percolation threshold found by Cohen {\it et
  al.}~\cite{cohen2000resilience}. Kenah and
Robins~\cite{kenah2007second} generalized the results in
Ref.~\cite{newman2002spread} and found that the SIR maps with a
semi-directed link percolation process and their theory predicts the
probability $\Pi$ of an epidemic in the thermodynamic
limit. Importantly, for a constant (homogeneous) recovery time, they
proved that at the final state, the value of $\Pi$ is equal to the
relative size of the giant component of link percolation,
$P_{\infty}$. However, for the case of non-constant (non-homogeneous)
recovery time and an infection time which follows an exponential
distribution, $\Pi<P_{\infty}$ for $T>T_c$. This implies that the
probability of an epidemic in the SIR model does not map to link
percolation, i.e., this percolation process cannot predict the
probability of an epidemic.

Several approaches have been developed to study the effects of the
community structure on the disease spreading. Salath\'e {\it et
  al.}~\cite{salathe2010dynamics} found that in networks with a strong
community structure, there exists a trapping effect because the
disease is more likely to stay inside the community than to reach
other communities. Besides, they obtained that such structure delays
the epidemic spreading across the network. Hindes {\it et
  al.}~\cite{hindes2013epidemic} presented the time evolution
equations in the thermodynamic limit for a network of sub-networks or
communities in which at a global scale each community is represented
by a ``supernode,'' and all supernodes are arranged in a 1-dimensional
lattice, i.e., each supernode has only two supernode neighbors. They
showed that if the disease starts in one of these communities, then the
intra-degree distribution affects the propagation front at a global
scale. Vazquez~\cite{vazquez2007epidemic} developed a model of
communities composed by a finite number of nodes and solved it
analytically. The author obtained that the disease spreading is
characterized by oscillations at the early stages of the dynamic, and
there exists a critical basic reproductive number (i.e., the number of
secondary cases from an index case) above which the disease reaches a
macroscopic number of communities. Colizza and
Vespignani~\cite{colizza2007invasion,colizza2008epidemic}, and
Barthelemy {\it et al.}~\cite{barthelemy2010fluctuation} studied
metapopulation systems or networks composed by sub-populations with
homogeneous mixing and homogeneous or heterogeneous degree between
sub-populations over which individuals diffuse. For this model, they
obtained deterministic reaction-diffusion equations that describe the
disease spreading and found that there is a global invasion threshold,
above which a pandemic emerges. They showed that this threshold
depends on the degree heterogeneity and the number of individuals or
agents moving among sub-populations. Recently, Sah {\it et
  al.}~\cite{sah2017unraveling} studied on several realistic social
networks of animals how the community structure affects disease
spreading. The database included a wide range of structures ranging
from quite homogeneous networks with a weak structure of communities
to networks with highly segregated or fragmented communities such as
raccoons, field voles, and northern elephant seals. The authors found,
based on simulations, that the community structure does not affect the
probability of epidemics or the fraction of infected nodes unless the
global network has a very strong or extreme community
structure. Finally, Nadini {\it et al.}~\cite{nadini2018epidemic}
studied temporal networks with communities in which the number of
nodes (size) in each community follows a power-law distribution. They
found that in the limit of highly segregated communities, the final
fraction of recovered nodes in the SIR model is reduced. Besides, in
this limit, they obtained that when the community size is
heterogeneous, the fraction of recovered nodes is higher than the case
of a constant community size.

While many of the studies mentioned above analyzed the effect of
communities on the disease spreading at a local and global scale, they
are based only on simulations, or they do not consider the internal
structure of the communities or sub-populations (such as in the case of metapopulation
networks). A theoretical model that predicts the probability
of a pandemic is still lacking in structured communities, that is,
communities with an internal static structure. 

In this paper, we develop a model and study it theoretically to
understand the disease spreading at a global scale for the case of a
static network with a strong community structure and find under which conditions a pandemic
occurs. Additionally, we study how this structure shapes the evolution
of the number of ``infected'' communities.  Finally, in contrast to
isolated networks~\cite{kenah2007second}, our work finds that link
percolation predicts the probability of a pandemic due to the weak
dependence of the global transmissibility (between communities) on their
global recovery time.

\section{Model}\label{sec.mod}
In this section, we explain the structure of the synthetic network
with communities and the disease spreading process. In this work, we
only consider static networks. Note that we
describe our model at two scales: 1) a meta-level or global scale in
which the communities are treated as supernodes and all the links
between any two communities are represented by a single superlink, and
2) a microscopic or local scale in which the process is described at
the level of the nodes and links in each community.

We consider a network of communities with a random structure in which
the nodes of each community have internal connectivity or degree $k$
that follows a distribution denoted as $P(k)$. The number of
communities is $N^g$, and the number of nodes in each community is
$N_i$ with $i=1,\dots,\;N^g$. For simplicity, we assume that all the
communities have the same internal degree distribution and the same
number of nodes $N_i=N$ that could be either finite or infinite.  The
bridge nodes in one community are the nodes with external links, i.e.,
that are connected to other communities~\cite{dong2018resilience}. In
our model, each bridge node always has only one link which connects to
another community.

\begin{figure}[H]
\vspace{0.5cm}
\begin{center}
\begin{overpic}[scale=0.5]{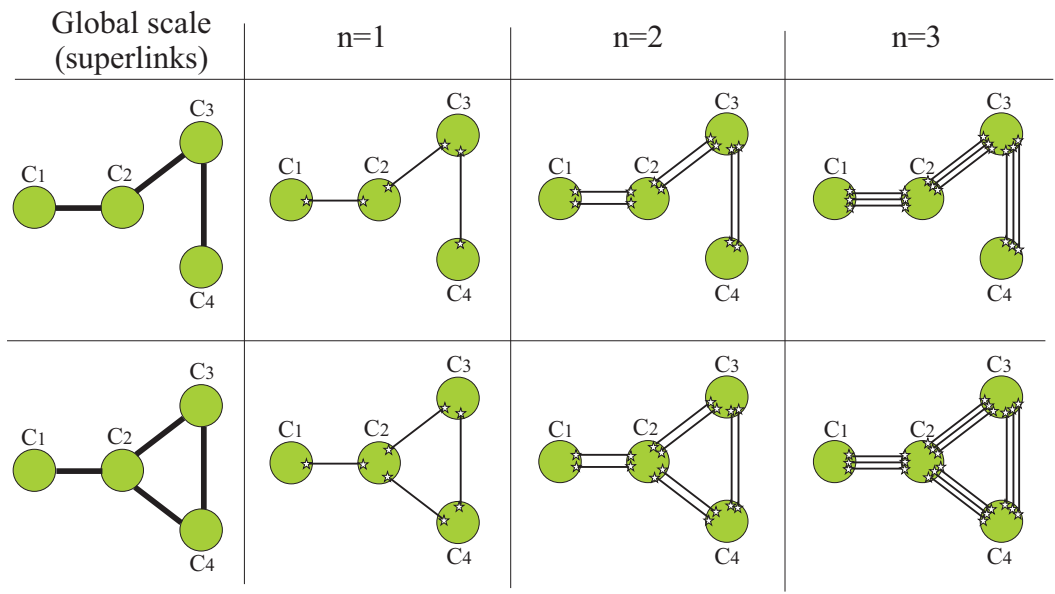}
  \put(0,25){}
\end{overpic}
\end{center}
\caption{Schematic illustration of four communities ($C_1$, $C_2$, $C_3$, and
  $C_4$, represented by circles) in which each row corresponds to a
  different global structure. The first column shows the global
  structure in which connections between communities are superlinks, and
  successive columns show different numbers of bridge nodes $n$ (stars)
  for the same global structure: $n=1$ (second column), $n=2$ (third
  column), $n=3$ (fourth column). In all configurations for the first
  row: $C_2$ and $C_3$ have two superlinks ($k^g=2$), while $C_1$ and $C_4$ have
  only one ($k^g=1$), while for the second row: $C_1$ has $k^g=1$, and
  $C_2$, $C_3$, and $C_4$ have $k^g=2$.}\label{fig.stars}
\end{figure}

When $n=1$, a community connects to another community only through one
bridge node (see Fig.~\ref{fig.stars}, second column). For $n>1$,
there are $n$ bridge nodes of $C_i$ that connect to $n$ bridge nodes
of $C_j$ (see Fig.~\ref{fig.stars}, third and fourth columns), where
$C_i$ and $C_j$ denote the communities $i$ and $j$, respectively. At a
global scale, all these links between $C_i$ and $C_j$ are represented
by a superlink, and we denote $P(k^g)$ as the fraction of communities
or supernodes with $k^g$ superlinks. Since a community with
$k^g<\infty$ superlinks has $nk^g<\infty$ external links, in the limit
$N\to \infty$ the number of external links for each community is
insignificantly smaller compared to the number of its internal
links. We refer to this structure or topology as an extreme or strong community
structure because the number $n$ of links between two communities is
finite and insignificant compared to the number of links inside each
community which is infinite in the thermodynamic limit. By increasing
$n$ we will show how a higher number of links among communities
induces a pandemic. To study the disease spreading at a global scale
from simulations, we consider that a community has an epidemic or a
supernode is ``infected'' if its number of infected nodes/individuals
is above a cutoff $s_c$, and susceptible if it is below $s_c$. Note
that the value of $s_c$ depends on the local degree distribution $P(k)$ of
each community and its number of nodes (in Appendix~\ref{ap.sc}, we
explain how to estimate $s_c$). The cutoff $s_c$ allows distinguishing
a macroscopic epidemic from a small outbreak.  After a community or
supernode is infected, it will go to the ``recovered'' state when all
the infected individuals within the community go to the recovered
state.

Using this network as a substrate, we study a discrete-time SIR
process. We define ``microscopic transmissibility'' $T$ as the effective
probability of infection between an infected node and its susceptible
neighbor.  At the microscopic level for the discrete-time dynamic, we
consider that a node infects a susceptible neighbor with probability
$\beta$ per unit time step, and it recovers after $t_r$ time steps. The
microscopic transmissibility for this model is given by,
\begin{eqnarray}\label{eq.defTT}
T&=&1-(1-\beta)^{t_r}.
\end{eqnarray}
Here, we will show the case of $t_r=1$, in which case $T=\beta$, but
qualitatively similar results are obtained for $t_r=5$.

\begin{figure}[H]
\vspace{0.5cm}
\begin{center}
\begin{overpic}[scale=0.6]{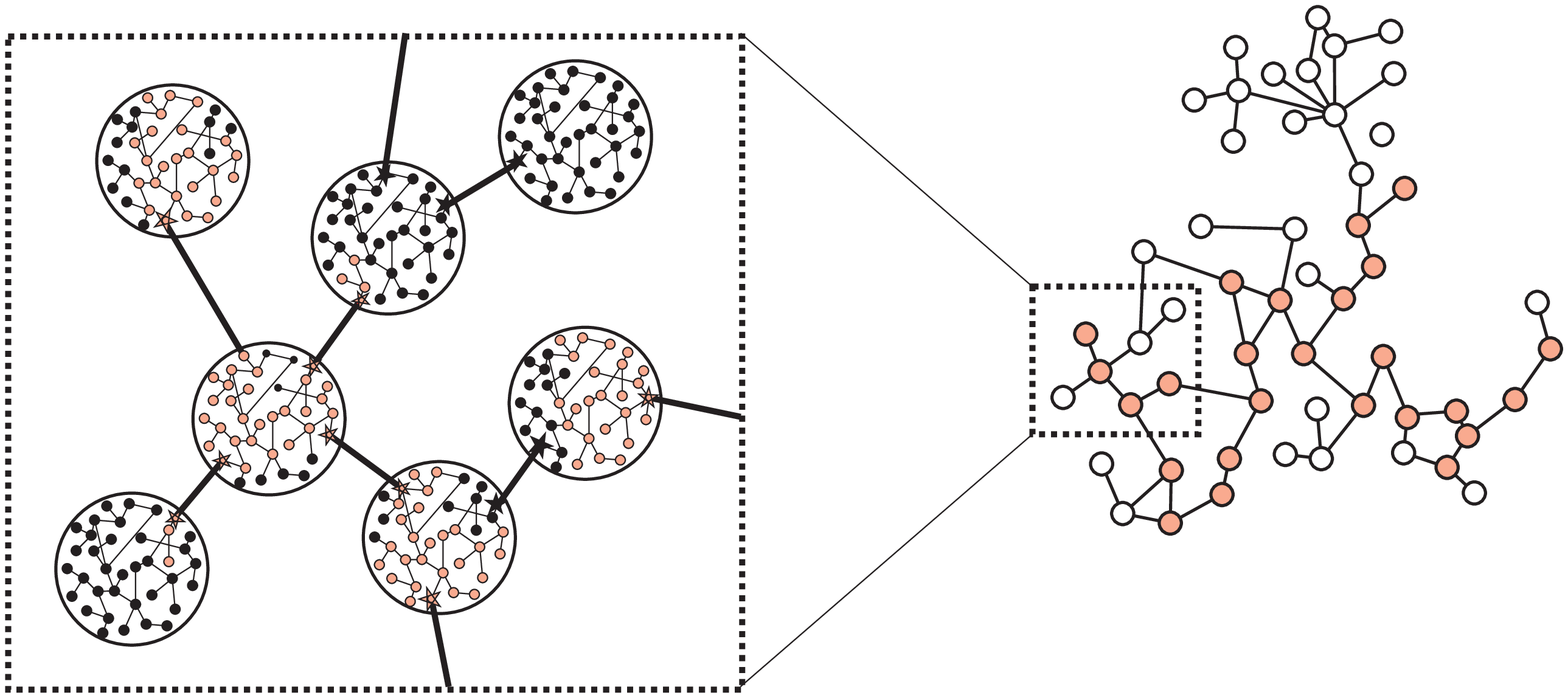}
  \put(45,50){}
\end{overpic}
\end{center}
\caption{Schematic illustration of the SIR model at the final state in a network
  with communities and $n=1$. On the left, each large circle represents
  a community, stars denote bridge nodes, and small circles are the
  internal nodes or individuals of each community. The pink nodes
  correspond to recovered individuals, while the black ones are
  susceptible. On the right, we show the network of communities at a
  global scale where each circle is a supernode or community. The pink
  supernodes are the communities where the epidemic developed, and are
  white otherwise, i.e., the disease did not reach the community or
  only developed as an outbreak. The area enclosed with a dotted line corresponds to
  the figure on the left.}\label{fig.esqglob}
\end{figure}

For the stochastic simulations in finite networks, at time $t=0$, all
the nodes of the whole network are susceptible except for one infected
node/individual in a community chosen at random. In our dynamic model,
we compute the temporal evolution of the fraction of infected
supernodes and the fraction of recovered supernodes at the final state
for a given value of $T$, which we denote as $I^g$ and $R^g$,
respectively. In finite networks, we consider that globally, the
disease turns into a pandemic if the number of infected supernodes at
the final state exceeds a threshold $s_c$. Note that at a global
scale, the value of $s_c$ depends on the global degree distribution
$P(k^g)$ of supernodes. In this paper, since the local and global
degree follow the same or similar degree distribution (for instance, a
power-law distribution with similar exponent values at a local and
global scale), we use the same value of $s_c$ to distinguish
outbreaks, epidemics, and pandemics (in Appendix~\ref{ap.sc}, we
explain the method to estimate $s_c$ based on simulations). On the
other hand, in the thermodynamic limit ($N\to \infty$ and $N^g\to
\infty$), a community/supernode has an epidemic if the fraction of
recovered individuals is not zero, while a pandemic takes place if the
fraction of recovered communities/supernodes is finite.
Fig.~\ref{fig.esqglob} shows a schematic illustration of the disease
spreading at microscopic and global scales. We define the ``global
transmissibility,'' $T^g$, as the effective probability of infection
between an infected supernode and its susceptible supernode neighbor.

In the following sections, we present a mathematical approach to compute the global
transmissibility and the relevant magnitudes that characterize the
disease spreading at a global scale based on simulations and theory.

\section{Microscopic and Macroscopic dynamic}\label{microMacrDy}
In this section, we study how the strong community structure affects
the epidemic spreading dynamic at a microscopic and macroscopic
scale. We denote an Erd\H{o}s R\'enyi network of Erd\H{o}s R\'enyi
communities as ER-ER, that is, $P(k)$ and $P(k^g)$ follow a Poisson
distribution, where $k$ and $k^g$ are the numbers of internal
connectivities of a node and the number of superlinks of a supernode,
respectively. Similarly, we denote a scale-free network of scale-free
communities as SF-SF, where the degree distributions decay as a
power-law with exponent $\lambda$. It is important to note that the
fraction of bridge nodes in any community is zero in the thermodynamic
limit because they have $n k^g<\infty$ bridge nodes. The internal
structure in each community is random, and at a global scale, the
communities or supernodes are also randomly connected through the
superlinks.

In the following, we show that the dynamic global spreading can be
described by an SIR model in an aggregated network in which the
supernodes do not have an internal structure, but they preserve the
same degree distribution of superlinks $P(k^g)$ as in the model where
supernodes are communities with internal structure.  Analogously to
the microscopic scale where the transmissibility is the
probability that an infected node transmits the disease to its
susceptible neighbor  (given by
Eq.~(\ref{eq.defTT})), in the aggregated network, we will obtain the global
transmissibility $T^g$  between an infected and susceptible supernodes. This
magnitude will be computed from the probability of recovery time
$P(\tau_R)$ and the distribution of infection time $P(\tau_I|\tau_R)$.
 Here, $\tau_R$ is the time between two
events in each community:
\begin{enumerate}
\item the moment at which the number of infected nodes in the community is above $s_c$
\item the moment when no more infected nodes exist in the community after the first event took place.
\end{enumerate}
The first event represents the fact that health authorities
declare an epidemic only after having a certain
number of infected individuals, and the second event represents the
moment at which the authorities declare that the community is free of the
epidemic.

\begin{figure}[H]
\vspace{0.0cm}
\begin{center}
\begin{overpic}[scale=0.30]{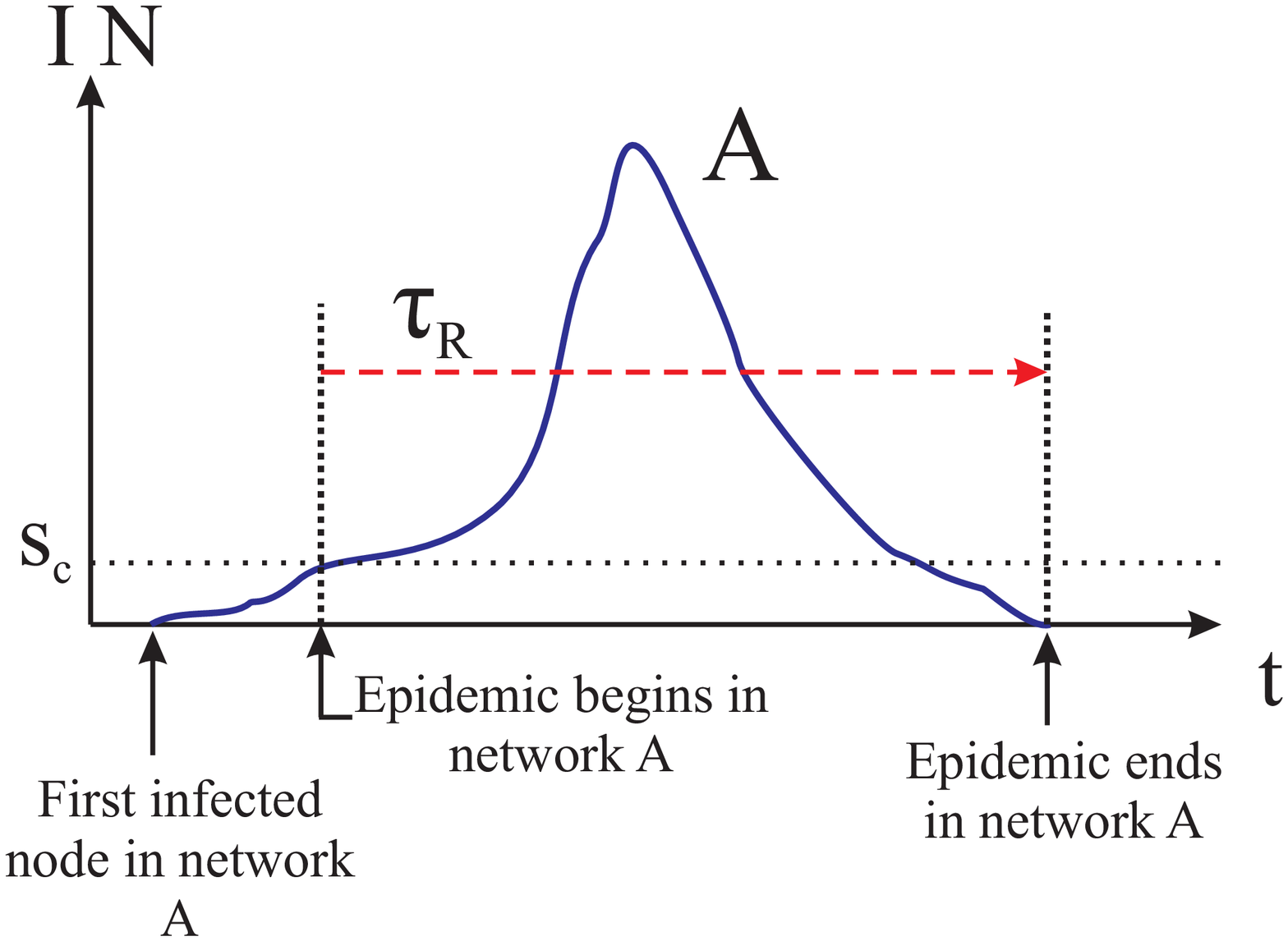}
  \put(65,70){(a)}
\end{overpic}
\begin{overpic}[scale=0.30]{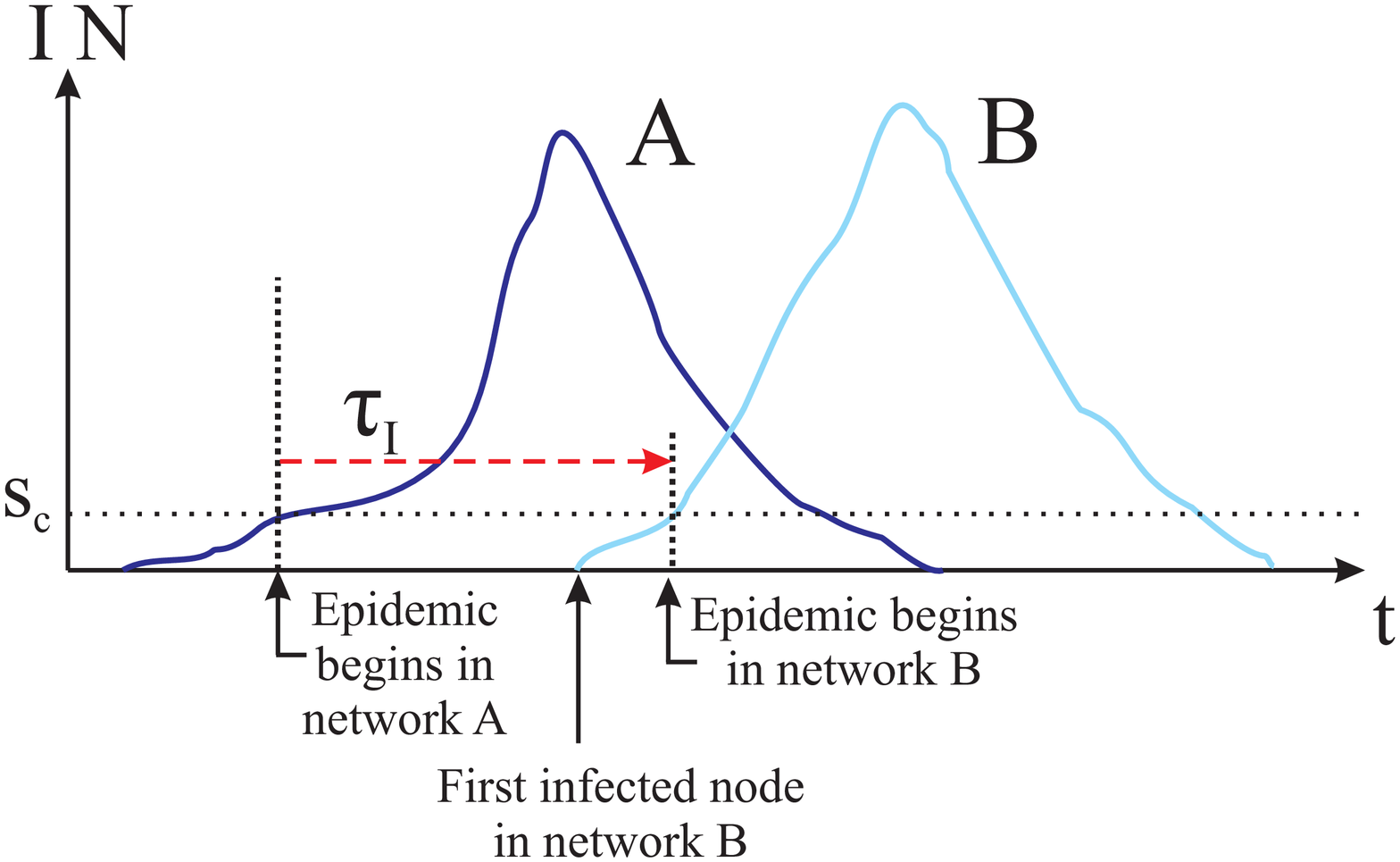}
  \put(75,58){(b)}
\end{overpic}
\end{center}
\caption{Schematic illustration  of the definition of $\tau_R$ (panel a) and
  $\tau_I$ when a community $A$ infects $B$ (panel b). The figures
  illustrate the time evolution of the number of infected nodes $I\times N$
  in community $A$ (dark blue) and community $B$ (light blue). The
  horizontal dotted line corresponds to the threshold $s_c$ above
  which a community is regarded as infected. In panel (a), the time $\tau_R$ (red dashed
  interval) corresponds to the time interval between the moment at
  which community $A$ becomes infected, and the moment it recovers. In
  panel (b), the time $\tau_I$ (red dashed interval) corresponds to the time
  interval between the times in which the two communities $A$ and $B$
  get infected. }\label{fig.tauitaiR}
\end{figure}

Similarly, $\tau_I$ is the period in which a community $A$ infects a
community $B$ (see schematic illustration in Fig.~\ref{fig.tauitaiR} of $\tau_I$
and $\tau_R$).  Using the above definitions of $\tau_I$ and $\tau_R$,
we define $T^g_{\tau_R}$ as the effective global transmissibility,
which is the conditional probability that a supernode with recovery
time $\tau_R$ infects its susceptible supernode neighbors and is given
by:
\begin{eqnarray}\label{eq.tgtaurr}
T^g_{\tau_R}&=&\sum_{\tau_I=0}^{\tau_R}P(\tau_I|\tau_R),
\end{eqnarray}
where $P(\tau_I|\tau_R)$ is the probability that a community $A$
infects another $B$ after $\tau_I$ time steps given that $A$ recovers after
$\tau_R$ time steps. We also define the total
effective global transmissibility $T^g$ as
\begin{eqnarray}
	T^g&=&\sum_{\tau_R=0}^{\infty}T^g_{\tau_R}P(\tau_R),
\end{eqnarray}
where $P(\tau_R)$ is the probability that a
community recovers after $\tau_R$ time steps since it was infected.

\begin{figure}[H]
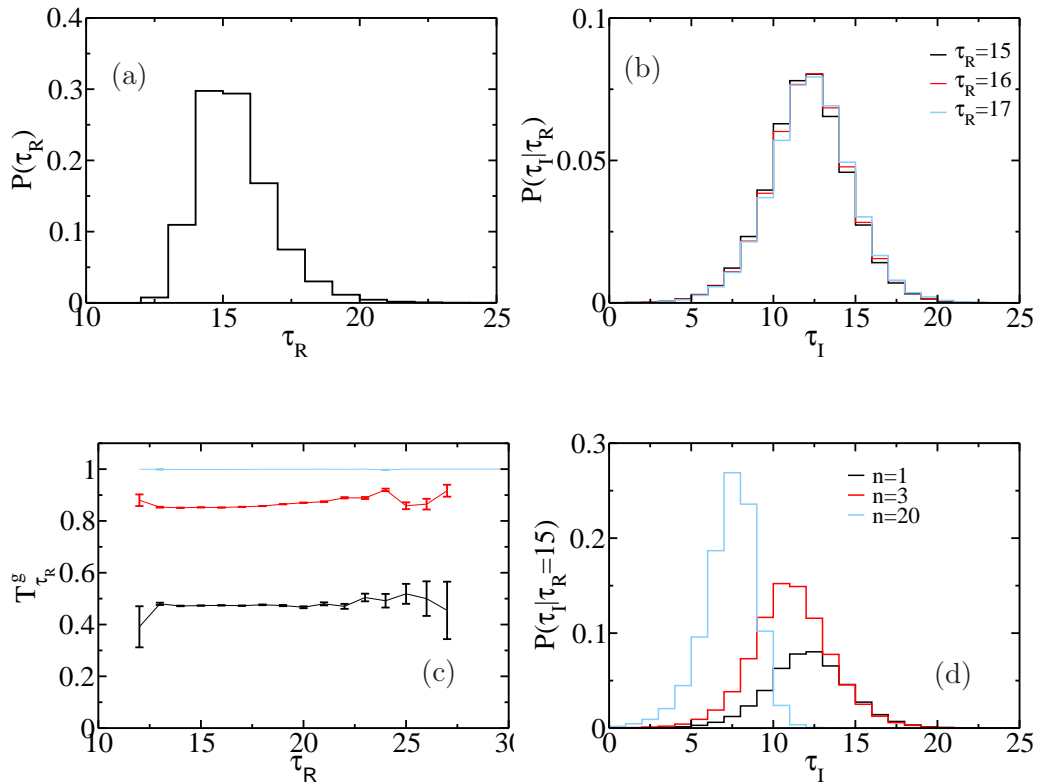

\vspace{0.5cm}
\begin{center}
\begin{overpic}[scale=0.25]{Fig04a.eps}
  \put(20,55){(a)}
\end{overpic}
\vspace{1cm}
\begin{overpic}[scale=0.25]{Fig04b.eps}
  \put(20,55){(b)}
\end{overpic}
\vspace{0.0cm}
\begin{overpic}[scale=0.25]{Fig04c.eps}
  \put(80,20){(c)}
\end{overpic}
\vspace{0.0cm}
\begin{overpic}[scale=0.25]{Fig04d.eps}
  \put(80,20){(d)}
\end{overpic}
\end{center}
\caption{Normalized time distribution of $\tau_R$ and $\tau_I$ for
  different values of $n$ for two connected ER communities with
  $\langle k \rangle =3$. Panel (a): distribution of $\tau_R$. Panel
  (b): the conditional distribution $P(\tau_I |\tau_R)$ for $n=1$
  and $\tau_R=15$ (black), $\tau_R=16$ (red), $\tau_R=17$
  (light blue). Panel (c): $T^g_{\tau_R}$ as a function of $\tau_R$ for
  $n=1$ (black), $n=3$ (red), and $n=20$ (light blue). Panel (d):
  the conditional distribution $P(\tau_I|\tau_R)$ for $\tau_R=15$ and $n=1$
  (black), $n=3$ (red), $n=20$ (light blue). The results were obtained with over
  $10^6$ realizations for $T=0.70$, $N=10^4$, and
  $s_c=100$.}\label{fig.hist}
\end{figure}

From our simulations on communities that are not aggregated (they have
internal structure), we obtain that the distribution of $\tau_R$ is
broad, as shown in Fig.~\ref{fig.hist}a. Therefore, although at a
microscopic level, the recovery time of individuals or nodes is unique
($t_r$ is the same for all nodes), the random internal structure of a
community induces a distribution of recovery time at a global
level. In Fig.~\ref{fig.hist}b, we show $P(\tau_I|\tau_R)$ for
different values of $\tau_R$ in which we observe that increasing
$\tau_R$ shifts slightly the probability $P(\tau_I|\tau_R)$ to the
right (larger $\tau_I$ values).

Due to our definitions of infected and recovered communities at a
global scale that are based on the cutoff $s_c$, we observe in
Fig.~\ref{fig.hist}b that for different values of recovery time
$\tau_R$, there is a range of values of $\tau_I$ in which
$\tau_I>\tau_R$. This behavior implies that after a community $A$ is
declared free of the epidemic, $A$ might infect a community $B$ which
seems to violate the causality of the spreading process because an
already recovered community/supernode cannot infect another
community/supernode. However, this case can occur since, at a
microscopic scale before $A$ recovers, its bridge nodes could transmit
the disease to community $B$. However, since the number of infected
nodes in $B$ is still below $s_c$ when $A$ recovers, at a global scale
$B$ is susceptible, which explains the problem with
causality. Nonetheless, because an increasing number of bridge nodes
does not change $P(\tau_R)$, but moves to the left the distribution
$P(\tau_I|\tau_R)$ (see Fig.~\ref{fig.hist}d), the probability of
$\tau_I>\tau_R$ decreases, thus the effect of lack of causality can be
disregarded. On the other hand, this shift also implies that $\tau_I$
could be negative, but the probability of such an event is very low
($P(\tau_I<0)\lesssim 10^{-4}$) for $n\leq20$.

Using the recovery and infection time distribution shown in
Fig.~\ref{fig.hist}a-b, we simulate a SIR model in a network with a
degree distribution given by $P(k^g)$. We set $P(\tau_I|\tau_R)=0$ for
$\tau_I>\tau_R$ and for $\tau_I<0$ to impose the causality.

In summary, to study the dynamic spreading for the aggregated network,
we follow the next four steps:
\begin{enumerate}
\item For a network with $n$ bridge nodes, with $N^g$ communities that follow a global
  degree distribution $P(k^g)$, and $N$ nodes in each community that
  follows a local degree distribution $P(k)$, we built a network with
  $N^g$ supernodes with the same degree distribution $P(k^g)$.
\item Given: i) the values of the probability of infection $\beta$ and the
  recovery time $t_r$, ii)  two communities with size $N$ and local
  degree distribution $P(k)$, and iii) these two communities have $n$ bridge
  nodes, we run the SIR model and compute the time distributions $P(\tau_I|\tau_R)$ and $P(\tau_R)$.
\item We run the SIR model in the aggregated network, using the
  distributions $P(\tau_I|\tau_R)$ and $P(\tau_R)$ as the infection
  time distribution between an infected and susceptible supernode, and
  the recovery time distribution of a supernode, respectively.
  \item We compute the fraction of infected supernodes $I^g$ and
    compare to the fraction of infected communities in the
    microscopic network using the same values of $\beta$ and $t_r$ in
    step 2.
\end{enumerate}

\begin{figure}[H]
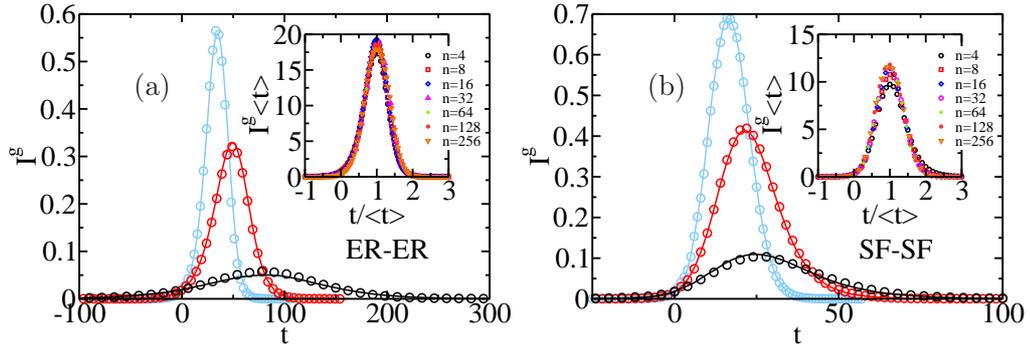

\vspace{0.5cm}
\begin{center}
\begin{overpic}[scale=0.25]{Fig05a.eps}
  \put(25,50){(a)}
\end{overpic}
\vspace{0.5cm}
\begin{overpic}[scale=0.25]{Fig05b.eps}
  \put(25,50){(b)}
\end{overpic}
\end{center}
\caption{Time evolution of the fraction of infected communities for
  different values of $n$: 1 (black), 3 (red), 20 (light blue). For each
  value of $n$, we show the average value of $I^g$ obtained from 100
  realizations of the aggregated network (symbols) and the network
  with communities (line).  Panel (a) corresponds to $T=0.70$ for an
  ER network composed of ER communities with $\langle k^g
  \rangle=\langle k \rangle =3$. Note that for $n=1$, the disease
  reaches a macroscopic fraction of communities only for $T\gtrsim
  0.6$ (see Fig.~\ref{fig.est}). Panel (b) corresponds to $T=0.60$ for
  SF networks at a global scale with $\lambda=3$ and $2\leq k^g \leq
  200$, with SF communities in which $\lambda=2.5$ and $2\leq k \leq
  200$. For the simulations, we use $N=10^4$, $N^g=5\times 10^3$, and
  $s_c=100$. We set the time $t=0$ as the moment at which $I^g
  N^g=s_c$~\cite{valdez2015predicting,miller2011edge}. The insets show
  $I^g\langle t\rangle$ as a function of $t/\langle t\rangle$ for
  different values of $n$, where $\langle t\rangle$ is the time at
  which the fraction of infected communities is maximum. These results
  were obtained from the aggregated network.}\label{fig.nomark}
\end{figure}

From Figs.~\ref{fig.nomark}a-b, we observe that $I^g$ obtained from
the SIR in the aggregated network is in very good agreement with the
results obtained from the SIR model on the network with communities,
in particular, when the number of bridge nodes $n$ increases. Similar
results are also obtained for other values of $T$ (see
Appendix~\ref{ap.addresults}), indicating that our model can be well
described at a global scale as an SIR model. We also obtain that the
area of $I^g$ as a function of $t$ is the same for different values of
$n$ (see insets in Fig.~\ref{fig.nomark}). In
Appendix~\ref{ap.results}, we show additional results on the effect of
$n$ on the average time $\langle t\rangle$ at which the fraction of
infected communities is maximum.

In the following section and Appendix~\ref{ap.Nondeg}, we show
that the probability of a pandemic at the final state is well
predicted by link percolation, although the
distribution of recovery times is non-homogeneous (which is in contrast with the results of
Ref.~\cite{kenah2007second}).

\section{Final state: general formalism and simulations}\label{sec.statGral}
\subsection{Theory and critical point for a pandemic}
Here we present the equations that describe the disease at the final
state using percolation theory and the generating function
formalism~\cite{newman2001random}.

Assuming that $n = 1$, if community $A$ develops an epidemic, the
effective or global probability of transmitting the epidemic to community $B$ depends
on the following events:
\begin{itemize}
 \item a bridge node in a community $A$ (that connects to community
   $B$) belongs to the GC of recovered nodes which size is above $s_c$. This event occurs
   with probability $R$.
 \item an infected bridge node transmits the disease to the bridge
   node in community $B$ with probability $T$ (see Eq.~(\ref{eq.defTT})).
 \item the disease in community $B$ becomes an
	 epidemic (i.e., $R>0$ in community $B$) with probability $\Pi$.
\end{itemize}

At a global scale, the
effective or global probability of infection from one community to
another is $T R\; \Pi$. Similarly, for the case of $n>1$
bridges, the effective probability of transmission is
\begin{eqnarray}\label{eq.Tefff}
1-(1-T R\; \Pi)^n\equiv T^g.
\end{eqnarray}
which is the probability that at least one bridge node in community
$A$ transmits the disease to a bridge node in community $B$ from which
an epidemic develops. Note that $R$ and $\Pi$ are magnitudes
relative to one community which depend on the microscopic
transmissibility $T$, and they are evaluated based on
Ref.~\cite{kenah2007second} (see a brief explanation in
Appendix~\ref{ap.Rvsr}).  For the case of a fixed or homogeneous
recovery time, $\Pi=R$~\cite{kenah2007second} which are obtained
solving the following equations:
\begin{eqnarray}
  f_{\infty}&=&1-G_1(1-T f_{\infty}),\label{eq.com11}\\
  R&=&1-G_0(1-Tf_{\infty}),\label{eq.com22}
\end{eqnarray}
where $f_{\infty}$ is the probability that a link leads to a
macroscopic recovered cluster of nodes in a branching process, and
$G_0(x)$ and $G_1(x)$ are the generating functions of the degree
distribution and the excess degree distribution of a node,
respectively~\cite{newman2002spread,braunstein2007optimal}.

Using the effective global transmissibility $T^g$, we compute the
fraction of recovered communities or supernodes $R^g$ at the final
state. This magnitude is obtained from two generating functions that
describe the network structure at a global scale, and are given by
\begin{eqnarray}
  G_0^g(x)&=&\sum_{k^g=0}^{\infty} P(k^g)x^{k^g},\label{eq.ge0}\\
  G_1^g(x)&=&\sum_{k^g=0}^{\infty} \frac{k^g P(k^g)}{\langle k^g \rangle} x^{k^g-1}.\label{eq.ge1}
\end{eqnarray}

With these generating functions $G_0^g(x)$ and $G_1^g(x)$, considering
the aggregated system as a single network in which nodes do not have any
internal structure, the equations of the SIR model at the final state
are given by
\begin{eqnarray}
  f_{\infty}^g&=&1-G_1^g(1-T^gf_{\infty}^g),\label{eq.Rg1}\\
  R^g&=&1-G_0^g(1-T^gf_{\infty}^g),\label{eq.Rg2}
\end{eqnarray}
where $T^g$ is the effective transmissibility between communities (see
Eq.~(\ref{eq.Tefff})), $R^g$ is the fraction of recovered communities,
and $f_{\infty}^g$ is the probability that a superlink leads to a
macroscopic recovered cluster of supernodes in a branching
process~\cite{newman2002spread,braunstein2007optimal}. Note that
Eqs.~(\ref{eq.Rg1}) and~(\ref{eq.Rg2}) are the same as the SIR model in
a network without
communities~\cite{newman2002spread,braunstein2007optimal} for a
transmissibility $T^g$. However, we are interested in understanding
how the microscopic transmissibility $T$ affects the order parameter
$R^g$ for a pandemic (replacing Eq.~(\ref{eq.Tefff}) in
Eqs.~(\ref{eq.Rg1})-(\ref{eq.Rg2}))~\footnote{This is since $T$ is the
 parameter of our microscopic model, which controls the basic
reproductive number~\cite{meyers2007contact}. This magnitude, $R_0$, is a
relevant measure in epidemiology to estimate the initial epidemic
growth.}.

Applying the technique used in
Refs.~\cite{newman2002spread,braunstein2007optimal} to find the
critical point, and using that $R=\Pi$ for a homogeneous recovery
distribution, we obtain from Eq.~(\ref{eq.Rg1}) and
Eqs.~(\ref{eq.com11})-(\ref{eq.com22}) that for ER-ER networks, the critical microscopic
transmissibility of a pandemic $T_{c,pand}$ (above which
$f_{\infty}^g>0$, that is, $R^g>0$) satisfies the following equation
\begin{eqnarray}\label{eq.tcpandd}
1-\left(\frac{\Delta}{T_{c,pand}} \right)^{1/2}&=&e^{-\langle k \rangle (T_{c,pand}\Delta )^{1/2}},
\end{eqnarray}
where $\Delta \equiv 1-(1-1/\langle k^g \rangle)^{1/n}$, $\langle k^g
\rangle$ is the mean number of the superlinks of each supernode, and
$\langle k \rangle$ is the mean connectivity inside each
community. Note that $T_{c,pand}$ depends on the number of bridge
nodes $n$, similar to the metapopulation
networks~\cite{colizza2007invasion,colizza2008epidemic,barthelemy2010fluctuation} where
the global invasion threshold depends on the number of individuals or
agents moving among sub-populations.

\begin{figure}[H]
\vspace{0.5cm}
\begin{center}
\begin{overpic}[scale=0.25]{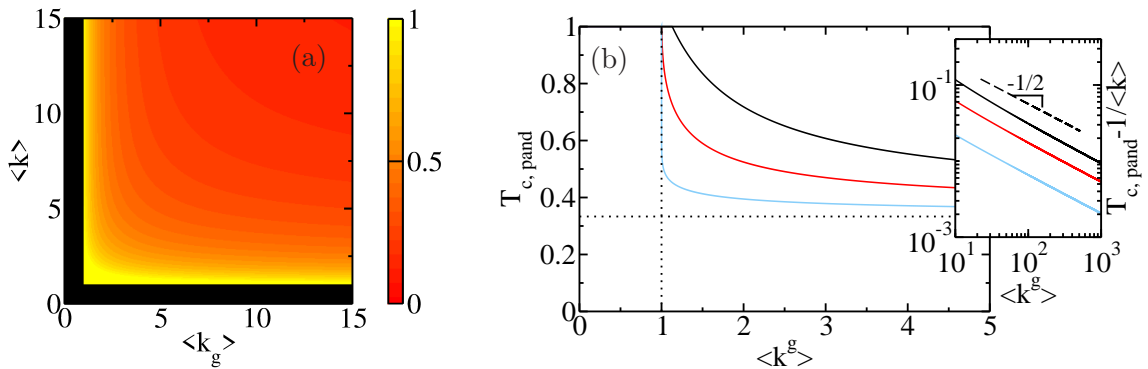}
  \put(65,70){(a)}
\end{overpic}
\hspace{0.5cm}
\begin{overpic}[scale=0.25]{Fig06b.eps}
  \put(14,47){(b)}
\end{overpic}
\end{center}
\caption{ Panel (a): Heat-map of the critical microscopic
  transmissibility for a pandemic, $T_{c,pand}$, in the plane $\langle
  k^g \rangle$ and $\langle k \rangle$ for an ER network of ER
  communities and $n=1$. The black region indicates that there is no
  pandemic phase in the network for any value of the microscopic
  transmissibility. Panel (b): Critical microscopic transmissibility
  for a pandemic $T_{c,pand}$ as a function of the global mean degree
  $\langle k^g \rangle$ for ER network of ER communities with $\langle
  k \rangle=3$ and different values of $n$: 1 (black), 3 (red), and 20
  (light blue). For each value of $n$, the system is in a pandemic phase
  above the curves, while below it is free of a pandemic. The vertical
  dotted line indicates the limit $\langle k^g \rangle=1$ and the
  horizontal dotted line corresponds to $T_{c,pand}=T_c=1/\langle k
  \rangle$. The inset shows $T_{c,pand}-1/\langle k \rangle$ as a
  function of $\langle k^g \rangle$ in log-log scale for the curves
  shown in the main plot. The curves and the surface are obtained from
  Eq.~(\ref{eq.tcpandd}). Note that the slope=-1/2 is predicted in
  Eq.~(\ref{eq.limitTc}). }\label{fig.T3d}
\end{figure}

From Eq.~(\ref{eq.tcpandd}), we obtain that $T_{c,pand}\to 1$ as
$\langle k\rangle$ and $\langle k^g\rangle$ decrease because in this
limit a pandemic only develops at the highest probability of
transmission to overcome the sparseness at a local and global
scale~\cite{cohen2003structural} (see Fig.~\ref{fig.T3d}a). On the
other hand, as $\langle k^g \rangle$ increases for a fixed value of
$\langle k \rangle$, $T_{c,pand}$ converges as a power-law to the
critical value of an isolated community $T_c=1/\langle k \rangle$ (see inset of
Fig.~\ref{fig.T3d}b). Expanding Eq.~(\ref{eq.tcpandd}) for $\langle
k^g \rangle \gg 1$, we obtain that $T_{c,pand}$ behaves as
\begin{eqnarray}\label{eq.limitTc}
T_{c,pand} \approx \frac{1}{\langle k
  \rangle}+\frac{1}{2}\left(\frac{1}{n\langle k^g \rangle
  \langle k \rangle}\right)^{1/2}.
\end{eqnarray}

From Eq.~(\ref{eq.limitTc}) we can see that $T_{c,pand}$ decreases
with the number of bridge nodes as a power-law, and for $n\to \infty$,
$T_{c,pand} \to T_c=1/\langle k\rangle$. Note that after an epidemic
develops in one community, the probability that the disease reaches
one bridge node increases with $n$. In turn, the probability that at
least one of the infected bridge node induces an epidemic in a
susceptible community also increases with $n$. As a consequence, for
large $n$, the disease cannot be confined in one community, and the
fluctuations of the early dynamic that extinguish the disease in a
community cannot ``halt'' the disease spreading at a global
scale. Therefore, in the limit $n\to \infty$ at the final state,
$T_c=T_{c,pand}$, and there is no distinction between the outcome of
an epidemic and pandemic since one implies the other (see
Eq.~(\ref{eq.limitTc})).

\subsection{Size and probability of a pandemic}
Besides the computation of the critical transmissibility for a
pandemic, it is also of interest to study the size of the pandemic in
terms of the number of recovered individuals and communities with
epidemics.

\begin{figure}[H]
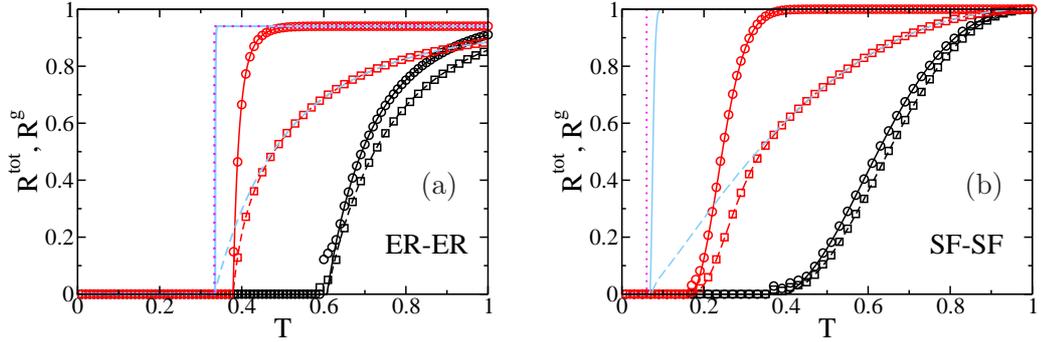

\vspace{0.5cm}
\begin{center}
\begin{overpic}[scale=0.25]{Fig07a.eps}
  \put(85,30){(a)}
\end{overpic}
\hspace{0.5cm}
\begin{overpic}[scale=0.25]{Fig07b.eps}
  \put(85,30){(b)}
\end{overpic}
\end{center}
\caption{Fraction of recovered individuals $R^{tot}$ ($\square$,
  dashed line) and communities that developed an epidemic $R^g$
  ($\bigcirc$, solid line) as a function of the microscopic
  transmissibility. Our results were obtained from the simulations
  (symbols) and Eqs.~(\ref{eq.Rg1})-(\ref{eq.Rg2}) (lines) for $n=1$
  (black), $n=20$ (red), and $n=10^4$ (light blue - only theory). The pink
  dotted line corresponds to the limit $n=\infty$ (see
  Eq.~(\ref{eq.stepf})). Panel (a) corresponds to an ER network of ER
  communities with $\langle k^g \rangle=\langle k \rangle =3$. Panel
  (b) corresponds to SF networks at a global scale with $\lambda=3$
  and $2\leq k^g \leq 200$, with SF communities in which $\lambda=2.5$
  and $2\leq k \leq 200$. The simulations were performed over 100
  network realizations with $N=10^4$, $N^g=5\times 10^3$, and (a)
  $s_c=600$ and (b) $s_c=100$.}\label{fig.est}
\end{figure}

In Fig.~\ref{fig.est}, we show the fraction of recovered individuals
$R^{tot}$ in the whole system and the fraction of communities that
developed an epidemic at the final state $R^g$ (where $R^{tot}\equiv
R^gR$) obtained from Eqs.~(\ref{eq.Rg1})-(\ref{eq.Rg2}) and
simulations. For ER-ER and SF-SF networks with $n=1$, there is little
difference between $R^{tot}$ and $R^g$ because the degree
distributions at a local ($P(k)$) and global ($P(k^g)$) scales are
similar, and $T^g\not\approx 1$ for $n=1$ (see
Eq.~(\ref{eq.Tefff})). However, as the number of bridge nodes
increases, the curves $R^{tot}$ and $R^g$ differ from each other,
particularly close to the critical point.

It is interesting to note that the fraction of recovered
individuals $R^{tot}$ converges to a function that vanishes
continuously at $T_{c,pand}$, in contrast to the fraction of recovered
communities $R^g$ that converges to a discontinuous step function for $n
\to \infty$:
\begin{eqnarray}\label{eq.stepf}
R^g(T,n=\infty) = \left\{%
\begin{array}{ll}
	\mbox{c}\;\; &\mbox{if}\;\;\; T>T_{c,pand}=T_c, \\
	0\;\;\; &\mbox{if }\;\;\; T\leq T_{c,pand}=T_c, 
\end{array}%
\right.
\end{eqnarray}
where $c>0$ and constant. This is because for any value of the
microscopic transmissibility $T>T_c$ when $n\to \infty$, the global
transmissibility tends to $T^g \to 1$ (see Eq.~(\ref{eq.Tefff})). In
consequence, if the epidemic begins in a community/supernode that
belongs to the GC of supernodes, the disease will reach all the
supernodes that belong to this cluster for any value of $T>T_c$. The value of
$R^g(T,n=\infty)$ is a constant and corresponds to the
fraction of supernodes that belong to the GC at a global scale. On the
other hand, for $T<T_c$ the disease never becomes an epidemic in a
community and hence it cannot becomes a pandemic which implies that
$R^g(T,n=\infty)=0$.

For the case $n=1$, the size of a pandemic is comparable (or
correlated) to the size of an epidemic in each community. In such a
scenario, if the authorities decide to apply a strong mitigation
strategy to prevent the disease spreading when the fraction of
infected communities is large, this also corresponds to a significant
fraction of infected individuals in each community. However, if the
number of bridge nodes $n$ increases, any strong response measure to halt an extended
global  disease could be considered ``disproportionate'' if
the size of the epidemic in each community is small, especially
near the critical point.  The increasing distance between the curves
$R^{tot}$ and $R^g$ as $n$ increases, establishes a problematic
scenario to any strategy that is based only on the number of infected
communities because if it is declared that a disease has reached a
pandemic status, it may be thought as alarmist since $R^{tot}\ll
R^g$. Therefore, this result suggests that the size of the epidemic in
each community could also be used to decide the required aggressiveness
of the mitigation strategy since this would allow identifying
pandemics that do not affect a substantial fraction of the population
near $T_c=T_{c,pand}$.

\begin{figure}[H]
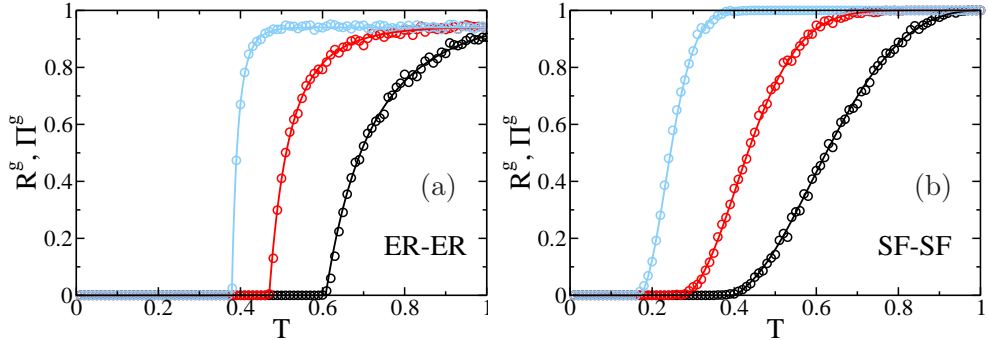

\vspace{0.5cm}
\begin{center}
\begin{overpic}[scale=0.25]{Fig08a.eps}
  \put(85,30){(a)}
\end{overpic}
\begin{overpic}[scale=0.25]{Fig08b.eps}
  \put(85,30){(b)}
\end{overpic}
\end{center}
\caption{Probability of a pandemic given that there is one community
  with an epidemic as a function of $T$ for $n=1$ (black), $n=3$
  (red), $n=20$ (light blue). Panel (a): the results correspond to an ER
  network of ER communities with $\langle k^g \rangle=\langle k
  \rangle =3$, and different values of $n$. Panel (b) corresponds to
  SF networks at a global scale with $\lambda=3$ and $2\leq k^g \leq
  200$, with SF communities in which $\lambda=2.5$ and $2\leq k \leq
  200$. The simulations were performed over $10^3$ network realizations
  with $N=10^4$, $N^g=5\times 10^3$, and (a) $s_c=600$ and (b) $s_c=100$. The
  lines ($R^g$) correspond to the theory  obtained from the
  Eqs.~(\ref{eq.Rg1})-(\ref{eq.Rg2}), and the symbols ($\Pi^g$) to the
  simulations.}\label{fig.probpand}
\end{figure}

Another significant concern for health authorities is the probability
of a false-positive pandemic alert because a false alarm would also
induce mistrust, panic, and fear in the population. In
Fig.~\ref{fig.probpand}a-b, we show the probability $\Pi^g$ that the
disease develops into a pandemic, given that there is at least one
community with an epidemic. Remarkably, we observe that this
probability is very close to the fraction of recovered communities at
the final state, i.e., $R^g \approx \Pi^g$, despite that the time
recovery distribution $P(\tau_R)$ is non-homogeneous (see
Fig.~\ref{fig.hist}a). This relation holds because the
transmissibility $T_{\tau_R}^g$ has a weak dependence on $\tau_R$ (see
Fig.~\ref{fig.hist}c). In fact, in Appendix~\ref{ap.Nondeg}, we show
based on a simple model that the SIR model with non-homogeneous
recovery time and constant $T_{\tau_R}^g$ maps into link percolation,
i.e., $R=\Pi$. Therefore our results in Fig.~\ref{fig.hist}c and
Appendix~\ref{ap.Nondeg} suggest that $R^g\approx \Pi^g$ in a network
with communities, and hence the probability of a pandemic converges to
the step function given in Eq.~(\ref{eq.stepf}).  Thus, after a
community develops an epidemic, not only a large number of communities
would develop epidemics (close to 100\%) if no  intervention from any authority is implemented, but also, it is very likely
to declare a pandemic. Besides, this implies that in a more
interconnected world and near the critical point, it is very likely
that health authorities will face a scenario in which the disease
reaches many regions (communities) with a small fraction of infected
individuals.

\section{Summary and conclusions}
In summary, we have studied the effect of extreme modularity in
structural modular networks on disease spreading at a global scale. We
found that the epidemic spreading through the network at a global
scale can be described as an SIR model with renormalized infection and
recovery distributions. On the other hand, as $n$ increases, the
probability and size of a pandemic increase and tend to a
discontinuous function of the transmissibility after the disease has
reached the status of the epidemic in one community. Besides, if the
transmissibility $T$ is close to the critical value of an epidemic,
our results indicate that the fraction of recovered communities is
significantly higher than the fraction of recovered individuals. This
situation can lead to a scenario in which a pandemic alarm could be
considered as an excessive alarm causing fear in the global
population. Finally, our simulations show that link percolation is a
good approximation to describe the final state of the disease
spreading at a global scale in random networks, although the recovery
time distribution of a community is non-homogeneous.

An important simplification of our work is that all communities have
the same degree distribution and the same number of nodes $N$ and
bridge nodes $n$. Our future studies will consider a distribution on these
magnitudes among the communities to explore how they affect the size
and probability of a pandemic.

\section{Acknowledgments}
 Boston University is supported by NSF Grants PHY-1505000, and by DTRA
 Grant HDTRA1-14-1-0017. LAB thanks UNMdP and CONICET (PIP 00443/2014)
 for financial support. S. H.  acknowledges financial support from the
 ISF, ONR, BSF-NSF: 2015781, ARO, the Israeli
 Ministry of Science, Technology and Space (MOST) in joint
 collaboration with the Japan Science Foundation (JSF), and the
 Italian Ministry of Foreign Affairs and International Cooperation
 (MAECI), and the Bar-Ilan University Center for Research in Applied
 Cryptography and Cyber Security.

\appendix

\section{Threshold $s_c$}\label{ap.sc}
\begin{figure}[H]
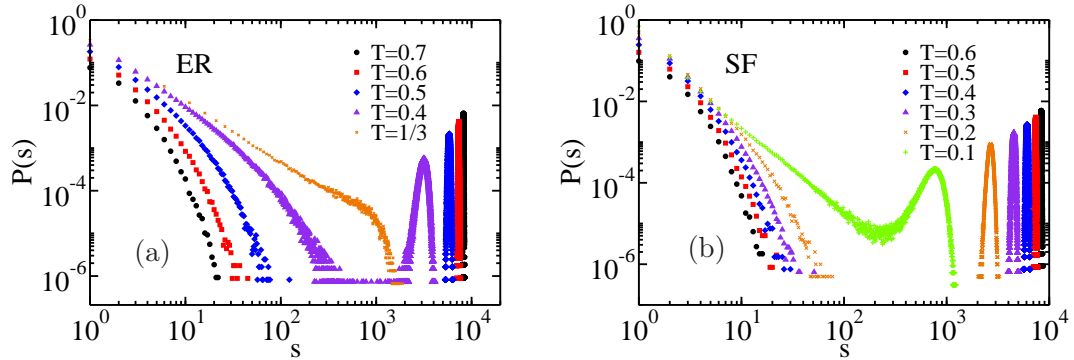

\vspace{0.5cm}
\begin{center}
\begin{overpic}[scale=0.25]{Fig09a.eps}
  \put(25,20){(a)}
\end{overpic}
\hspace{0.5cm}
\begin{overpic}[scale=0.25]{Fig09b.eps}
  \put(25,20){(b)}
\end{overpic}
\end{center}
\caption{Distribution of the number of recovered nodes at the final
  state, $P(s)$ for different values of $T$ for networks without
  community structure. Panel (a) corresponds to an ER network with
  $\langle k \rangle=3$. Panel (b) corresponds to a SF network with
  $\lambda=2.5$, $k_{min}=2$, and $k_{max}=200$. The simulations
  results were averaged over $10^6$ network
  realizations with $N=10^4$.}\label{fig.ap.PS}
\end{figure}

In the simulations of the SIR model,
fluctuations due to stochasticity could lead that the number of
infected nodes vanishes fast after the disease spreading started and
the number of recovered nodes is very small compared to the size of the
system, even for high values of transmissibility close to
$T=1$. Lagorio {\it et al.}~\cite{lagorio2009effects} proposed a
method to distinguish an outbreak from an epidemic, computing the
distribution of final sizes $P(s)$ of the disease from the simulations
of the SIR model. For $T>T_c$, $P(s)$ has a bimodal behavior, as shown
in Fig.~\ref{fig.ap.PS}. The left side of the distribution corresponds
to outbreaks, while the peak on the right corresponds to
epidemics. Between these two regions, there is a gap in which the
probability $P(s)$ is null. Therefore, any value of the threshold
$s_c$ that belongs to this region can be used to distinguish epidemics
and outbreaks. In Figs.~\ref{fig.ap.PS}a-b, we observe that as the
transmissibility $T$ approaches $T_c$ from above, that gap mentioned
above shrinks, and the distribution corresponding to outbreaks becomes
broader, and hence the minimum possible value of $s_c$ increases.

From the distributions $P(s)$ in Figs.~\ref{fig.ap.PS}a-b, we estimate the
values of $s_c$ that we use in this research. In the main text,
Sec.~\ref{microMacrDy} we choose $s_c=100$ for ER networks with
$\langle k\rangle=3$ and $T=0.7$, and $s_c=100$ for SF networks at
$T=0.60$, since this threshold distinguish outbreaks and epidemics. On
the other hand, to explore the size and probability of a
pandemic for different values of $T$, we set:
\begin{itemize}
\item $s_c=600$ for ER networks with $\langle k\rangle =3$ which is
  a sufficient threshold to distinguish epidemic for $T>0.4$. Note
  that $T_c=1/3$.
\item $s_c=100$ for SF networks, which is a sufficient threshold to
  distinguish an epidemic for $T>0.2$.
\end{itemize}

\section{Results for different values of $T$ and $n$}\label{ap.results}
\begin{figure}[H]
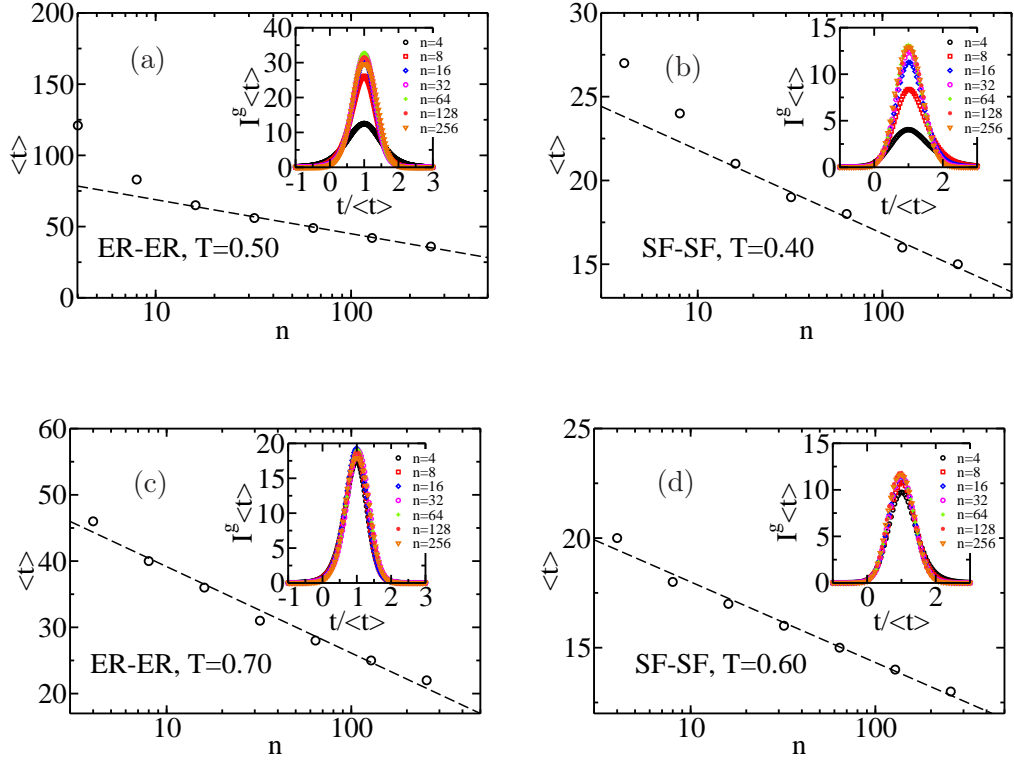

\vspace{0.5cm}
\begin{center}
\begin{overpic}[scale=0.25]{Fig10a.eps}
  \put(25,57){(a)}
\end{overpic}
\hspace{0.5cm}
\vspace{1cm}
\begin{overpic}[scale=0.25]{Fig10b.eps}
  \put(25,57){(b)}
\end{overpic}
\begin{overpic}[scale=0.25]{Fig10c.eps}
  \put(25,57){(c)}
\end{overpic}
\hspace{0.5cm}
\begin{overpic}[scale=0.25]{Fig10d.eps}
  \put(25,57){(d)}
\end{overpic}
\end{center}
\caption{The time $\langle t \rangle$ at which the fraction of
  infected communities is maximum, as a function of $n$ in linear-log
  scale. Panel (a) and (c) corresponds to $T=0.50$ and $T=0.70$,
  respectively, for an ER network of ER communities with $\langle k^g
  \rangle=\langle k \rangle =3$. Panel (b) and (d) corresponds to
  $T=0.40$ and $T=0.60$, respectively, for SF networks at a global
  scale with $\lambda=3$ and $2\leq k^g \leq 200$, with SF communities
  in which $\lambda=2.5$ and $2\leq k \leq 200$. The dashed line
  corresponds to a logarithmic fit $\langle t\rangle =A+B\ln(n)$,
  where: $A=92.8$ and $B=-10.4$ (panel a), $A=26.4$ and $B=-2.16$ (panel
  b), $A=52.2$ and $B=-5.7$ (panel c), and $A=21.7$ and $B=-1.6$ (panel
  d). For the simulations, we use $N=10^4$, $N^g=5\times 10^3$, and
  $s_c=100$. We set the time $t=0$ as the moment at which $I^g
  N^g=s_c$. The insets show $I^g\langle t\rangle$ as a function of
  $t/\langle t\rangle$ for different values of
  $n$.}\label{fig.appntmed}
\end{figure}

In Fig.~\ref{fig.appntmed} we show the time $\langle t \rangle$ at
which $I^g$ is maximum as a function of $n$ for different topologies
and values of the transmissibility. We observe that for large values of $n$,
$\langle t \rangle$ behaves as a logarithm function. Besides, we
obtain that the area of $I^g$ as a function of $t$ converges to the same value as $n$
increases (see insets in Fig.~\ref{fig.appntmed}).

\section{Macroscopic dynamic: additional results }\label{ap.addresults}
\begin{figure}[H]
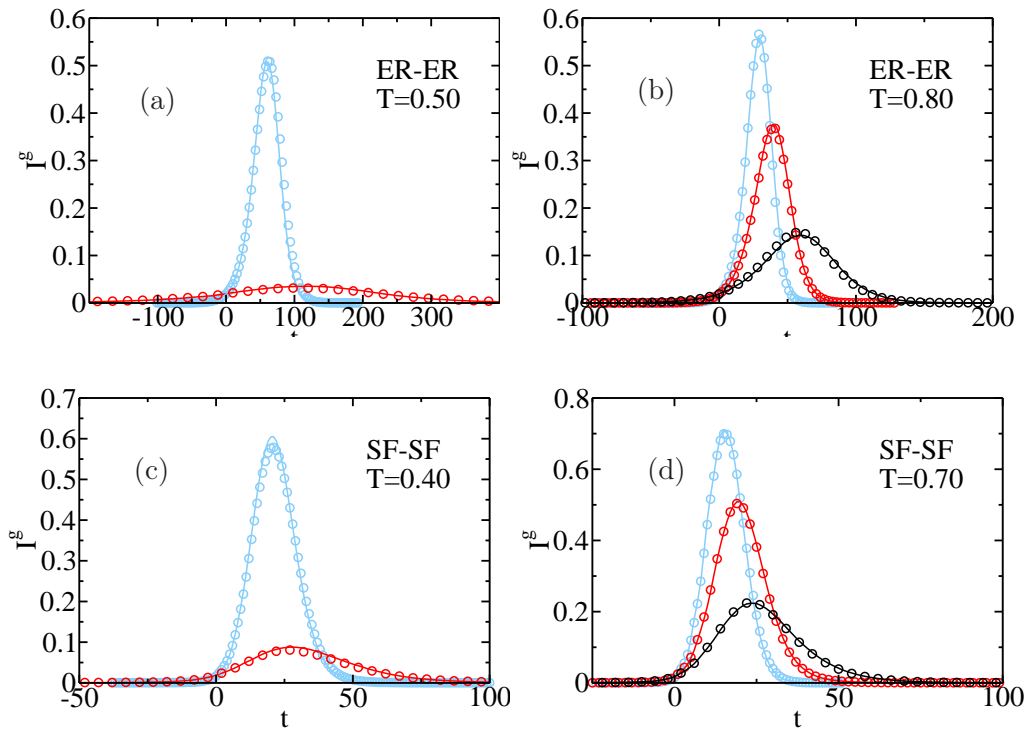

\vspace{0.5cm}
\begin{center}
\begin{overpic}[scale=0.25]{Fig11a.eps}
  \put(25,50){(a)}
\end{overpic}
\vspace{0.5cm}
\begin{overpic}[scale=0.25]{Fig11b.eps}
  \put(25,50){(b)}
\end{overpic}
\vspace{0.5cm}
\begin{overpic}[scale=0.25]{Fig11c.eps}
  \put(25,50){(c)}
\end{overpic}
\begin{overpic}[scale=0.25]{Fig11d.eps}
  \put(25,50){(d)}
\end{overpic}
\end{center}
\caption{Time evolution of the fraction of infected communities for
  different values of $n$: 1 (black), 3 (red), 20 (light blue). Other
  parameters are the same as in Figs.~\ref{fig.nomark}a-b. For each
  value of $n$, we show the average value of $I^g$ obtained from 100
  realizations of the aggregated network (symbols) and the network
  with communities (line).  Note that for panel (a) and (c), we do not
  show $I^g$ for $n=1$ because, in that case, the transmissibility is
  close or below $T_{c,pand}$ (see
  Figs.~\ref{fig.est}a-b).}\label{fig.nomarkaddresults}
\end{figure}

In Figs.~\ref{fig.nomarkaddresults}a-b, similar to Figs.~\ref{fig.nomark}a-b, we show
the time evolution of $I^g$ obtained from the aggregated network and
from the microscopic model for other values of transmissibility
$T$. In all cases, we observe an agreement between the results for the
aggregated network and the microscopic model.

\section{Percolation in semi-directed networks and the SIR model}\label{ap.Rvsr}
In this appendix, we review
Refs.~\cite{kenah2007second,kenah2011epidemic} in the first two
sections, which showed the mapping between SIR and a percolation
process in a semi-directed network. In the third section, we develop a
simple model which shows that the SIR model with non-homogeneous
recovery time maps into link percolation if $T_{\tau_R}^g$ is
constant.
\subsection{Method}\label{ap.RvsrMethod}
The key idea to study the final state of the SIR model using
percolation theory is to consider that there is a mapping between the
set of realizations of the stochastic SIR simulations and a
percolation process in a semi-directed network. To see this mapping,
let us consider that during the stochastic simulation of the SIR
model, node $i$ is infected at time $t$. Immediately after that, the
algorithm of the simulation generates a random recovery time $\tau_R$
obtained from a probability distribution $P(\tau_R)$. Hence, node $i$ will
recover at time $t+\tau_R$. Similarly as in the Gillespie algorithm,
for each neighbor of $i$, a random time $\tau_I$ is generated 
following a distribution $P(\tau_I|\tau_R)$ in which node $i$ transmits the
disease (since the moment that $i$ was infected).

Alternatively, instead of this procedure generating random numbers
``on the fly,'' i.e., during the simulation of the dynamic process,
the random numbers $\tau_R$ and $\tau_I$ can be obtained before
starting the dynamic. More specifically, a recovery time $\tau_R$ is
generated from a distribution $P(\tau_R)$ for each node before an
index case appears in the network. Note that the generation of
$\tau_R$ does not guarantee that a node $i$ will be infected, but in
case node $i$ gets infected during the dynamic process (that we
explained below), it would recover after a period $\tau_R$. After we
obtain the value of $\tau_R$ for a node $i$, we generate the times the
disease will take to reach each neighbor $j$ of $i$, including the
possibility that $\tau_I=\infty$, in which case, node $i$ will never
infect node $j$. Each link from $i$ with $\tau_I<\infty $ is
represented by an occupied arrow from $i$ to the other node connected
through this link. Analogously to the case of the recovery time, an
arrow from $i$ to $j$ does not mean that $i$ will effectively infect
$j$, but in case $i$ gets infected at time $t$ during the dynamic,
then $j$ would be infected at time $t+\tau_I$ (if another node does
not infect $j$ before this time).

The process described above does not develop the dynamic but only
generates all random numbers $\tau_R$ and $\tau_I$ before starting the
dynamic. In the case where two nodes point to each other, their link
is occupied and undirected, and if there is no arrow between these
nodes, their link is unoccupied. As a result of this procedure, we
obtain a semi-directed network.

After assigning all the times $\tau_R$ and $\tau_I$, a random node is
chosen as the index case, and then, the dynamic of the disease
spreading consists in following the arrows that emerge from the index
case, as described in
Refs.~\cite{kenah2007second,kenah2011epidemic}. If another node is
chosen as the index case, the branch of infection would be different,
so the semi-directed network contains many realizations of the SIR
model. Although this process is an alternative approach to ``on the
fly'' algorithm, it also allows interpreting many realizations of the
SIR model as a semi-directed network. We will see below that this
interpretation is useful for calculating the probability of an
epidemic $\Pi$ and the fraction of recovered nodes $R$
at the final state.

\subsection{Relationship between the in-component and out-component with $R$ and $\Pi$ }

In this section, we introduce several definitions of semi-directed
networks and then their relation to the fraction of recovered nodes
and the probability of an epidemic.

In any  semi-directed network, each node $i$ has three types of degree or connections:
\begin{itemize}
\item indegree: the number of incoming links to $i$,
\item outdegree: the number of outgoing links from $i$,
\item undirected degree: the number of undirected links of $i$.
\end{itemize}
The generating function of the probability $p_{abc}$ that a node has indegree ``$a$,''
outdegree ``$b$,'' and undirected degree ``$c$'' is given by
\begin{eqnarray}\label{ap.eq.Gxyu}
 G_0(x,y,u)&=& \sum_{a=0}^{\infty}\sum_{b=0}^{\infty}\sum_{c=0}^{\infty}p_{abc}x^{a}y^b u^c,
\end{eqnarray}

The mean indegree $\langle k_{in}\rangle$, outdegree $\langle
k_{out}\rangle$, and undirected degree $\langle k_{u}\rangle$ are
\begin{eqnarray}
  \langle k_{in}\rangle&=&\frac{\partial G_0}{\partial x}(1,1,1)=\sum_{a=0}^{\infty}\sum_{b=0}^{\infty}\sum_{c=0}^{\infty}a\;p_{abc},\label{eq.Ap.kkin}\\
  \langle k_{out}\rangle&=&\frac{\partial G_0}{\partial y}(1,1,1)=\sum_{a=0}^{\infty}\sum_{b=0}^{\infty}\sum_{c=0}^{\infty}b\;p_{abc},\\
  \langle k_{u}\rangle&=&\frac{\partial G_0}{\partial u}(1,1,1)=\sum_{a=0}^{\infty}\sum_{b=0}^{\infty}\sum_{c=0}^{\infty}c\;p_{abc}.
\end{eqnarray}
Since the total number of incoming connections is the same as the
total number of outgoing connections, then $\langle k_{in} \rangle
=\langle k_{out} \rangle \equiv \langle k_d \rangle$.

In a branching process, if we choose a node through:  a directed link following its
direction (forward), a directed link
going in the opposite direction (reverse or backward), or through a link without direction (undirected),
the generating functions that the reached node has indegree ``$a$'',
outdegree ``$b$'', and undirected degree ``$c$'' are given by
\begin{eqnarray}
 G_f(x,y,u) &=& \frac{1}{\langle k_d \rangle}\frac{\partial G_0}{\partial x}(x,y,u),\\
 G_{r}(x,y,u) &=& \frac{1}{\langle k_d \rangle}\frac{\partial G_0}{\partial y}(x,y,u),\\
 G_u(x,y,u) &=& \frac{1}{\langle k_u \rangle}\frac{\partial G_0}{\partial u}(x,y,u),
\end{eqnarray}
respectively.

Following the definitions of Ref.~\cite{newman2001random}, in semi-directed
networks, there exist for each node $i$ an:
\begin{itemize}
\item in-component that is the set of nodes from which $i$ can be
  reached by following the arrows. We define this component as
  macroscopic in-component ($\mathcal{M}_{in}$) if the number of nodes
  of this set is macroscopic. Otherwise, it belongs to a finite
  in-component.
\item out-component that is the set of nodes that can be reached from
  $i$ following the arrows. We define this component as
  macroscopic out-component ($\mathcal{M}_{out}$) if the number of nodes
  of this set is macroscopic. Otherwise, it belongs to a finite out-component.
\end{itemize}

\begin{figure}[H]
\vspace{0.5cm}
\begin{center}
\begin{overpic}[scale=1.3]{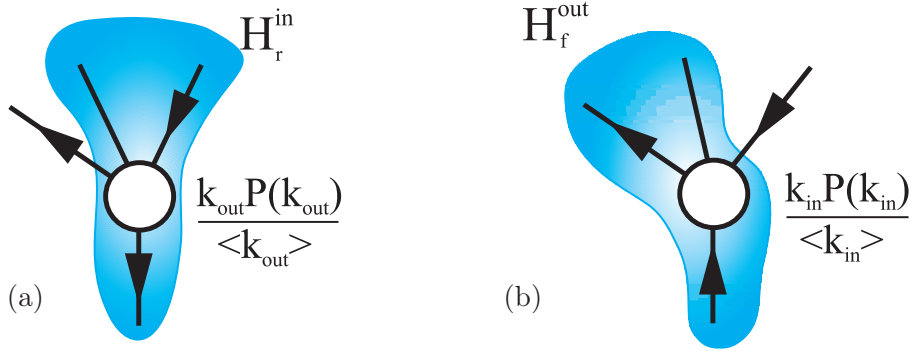}
  \put(0,5){(a)}
  \put(55,5){(b)}
\end{overpic}
\end{center}
\caption{Schematic figure of the backward (panel a) and forward
  branching (panel b). The direction of the branching process goes
  from bottom to top. Solid lines without any arrow represent undirected links,
  and arrows represent links with a direction. The blue area
  depicts the set of links used in the backward branching
  (panel a) and forward branching (panel b). For a backward (forward)
  branching, a node is reached through one of its outgoing
  (incoming) links with probability $k_{out}P(k_{out})/\langle k_{out}
  \rangle$ ($k_{in}P(k_{in})/\langle k_{in} \rangle$) and the
  in-component (out-component) continue to grow through its incoming
  (outgoing) and undirected links.}\label{fig.ap.HH}
\end{figure}

For a randomly chosen node, the generating functions of the {\bf
  finite} sizes of its in-component and out-component are denoted by
$H^{in}(z)$ and $H^{out}(z)$, respectively. These generating functions
can be obtained using a backward and forward branching, that is,
following the arrows in the opposite and along to their directions,
respectively.  Note that in this branching process, it is assumed that
the network is in the thermodynamic limit ($N\to \infty $) and the
structure is random. For a backward branching process, the generating
functions of the size of an in-component corresponding to a randomly
chosen node through a link, are
\begin{itemize}
\item $H_r^{in}(z)$ if the node is reached when going in the opposite direction of an arrow (see Fig.~\ref{fig.ap.HH}a) 
\item $H_u^{in}(z)$ if the node is reached through an undirected link.
\end{itemize}
Note that we do not consider $H_{f}^{in}(z)$, i.e., when a node is
reached following the direction of an arrow because, in this case, the
branching process would not correspond to an in-component.  Analogously,
for a forward branching process, the generating functions of the size
of an out-component corresponding to a randomly chosen node through a
link, are
\begin{itemize}
\item $H_{f}^{out}(z)$ if the node is reached going in the same direction of an arrow (see Fig.~\ref{fig.ap.HH}b) 
\item $H_u^{out}(z)$ if the node is reached through an undirected link.
\end{itemize}

For the case of a semi-directed network constructed by the procedure
explained in the previous section, if the index case has a finite
out-component, the disease can only reach a finite number of nodes
following the arrows. Therefore, the probability that an index case
does not trigger an epidemic, $1-\Pi$, is equal to the probability that
it belongs to a finite out-component $H^{out}(1)$. Otherwise, if the
index case belongs to the $\mathcal{M}_{out}$, this realization of the
SIR model corresponds to an epidemic. Besides,
Ref.~\cite{kenah2007second} also showed that using the same
semi-directed network, the fraction of recovered nodes $R$ is equal to
the probability that a node belongs to an $\mathcal{M}_{in}$
($R=1-H^{in}(1)$). To see this, let us assume that there is an
infinitesimal but not null fraction $\epsilon $ of infected nodes
during the dynamic spreading in the semi-directed network, in which
case there is an epidemic (i.e., $R\geq\epsilon>0$)~\footnote{Setting
  that $\epsilon \neq 0$ means that the computation of $R$ assumes
  that there is an epidemic.}. In the case for any susceptible node
$i$ that has an $\mathcal{M}_{in}$, at least one of the nodes in its
$\mathcal{M}_{in}$ will be infected in the thermodynamic limit (with
probability 1). Consequently, the disease will reach node $i$
following the arrows of the semi-directed network. However, if a
susceptible node $i$ has a finite in-component in which all of its
nodes are susceptible, then $i$ will never be reached by the
disease. In turn, the probability that at least one of the nodes of
this finite in-component is infected, vanishes as $\epsilon \to 0$, and
hence the disease can only reach the nodes within a
$\mathcal{M}_{in}$. Therefore, when there is an epidemic, the fraction
of nodes within a $\mathcal{M}_{in}$ is equal to the fraction of
recovered nodes $R$ at the final state.

In the following, we present the explicit relation between the
generating functions of the degree of a semi-directed network and the
SIR model described in Sec.~\ref{ap.RvsrMethod}.

For a node $i$ with a recovery time $\tau_R$ (with probability
$P(\tau_R)$), the probability that each connection is:
\begin{itemize}
\item  occupied and outgoing  is $T_{\tau_R}(1-T)$, i.e., node $i$ points its neighbor, but its neighbor does not point to $i$,
\item  occupied and incoming is $(1-T_{\tau_R})T$, i.e., $i$ does not point to its neighbor, but its neighbor points $i$,
\item  occupied and undirected is $T_{\tau_R}T$,
\item  unoccupied is $(1-T_{\tau_R})(1-T)$.
\end{itemize}
where $T_{\tau_R}=\sum_{\tau_I=0}^{\tau_R}P(\tau_I|\tau_R)$ is the transmissibility given that
node $i$ has recovery time $\tau_R$, and
$T=\sum_{\tau_R=0}^{\infty}T_{\tau_R}P(\tau_R)$ is the total
transmissibility.

Since the generating function of the total degree
of a node is $G_0(z)=\sum P(k) z^k$, then the generating function of
the probability $p_{abc}$ that a node has indegree ``$a$'', outdegree ``$b$'',
and undirected degree ``$c$'' (see Eq.~(\ref{ap.eq.Gxyu})) can be
rewritten as
\begin{eqnarray}
  G_0(x,y,u)&=&\sum_{k=0}^{\infty}P(k)\sum_{\tau_R=0}^{\infty}P(\tau_R)\times\notag\\
  &&[(1-T_{\tau_R})(1-T)+(1-T_{\tau_R})Tx+T_{\tau_R}(1-T)y+T_{\tau_R}Tu]^k.
\end{eqnarray}

Following Ref.~\cite{kenah2007second}, the generating function $H^{out}(z)$ is
obtained from the following equations
\begin{eqnarray}
  H_{f}^{out}(z)&=&zG_f(1,H_{f}^{out}(z),H_u^{out}(z)),\label{eq.Ap.Hfout}\\
  H_u^{out}(z)&=&zG_u(1,H_{f}^{out}(z),H_u^{out}(z)),\\
  H^{out}(z)&=&zG_0(1,H_{f}^{out}(z),H_u^{out}(z)).
\end{eqnarray}

For the case of homogeneous recovery time ($\tau_R$ is constant), these equations are reduced to
those proposed by Newman~\cite{newman2002spread} using an analogy between the SIR
model and link percolation:
\begin{eqnarray}
  f_{\infty}&=&1-G_1(1-Tf_{\infty}),\label{eq.ap.finf}\\
  R&=&1-G_0(1-Tf_{\infty}).\label{eq.ap.repid}
\end{eqnarray}
where $f_{\infty}$
is the probability that a link leads to a macroscopic recovered
cluster of nodes in a branching
process~\cite{newman2002spread,braunstein2007optimal}.

On the other hand, the generating function $H^{in}(z)$ is
obtained from the following equations
\begin{eqnarray}
  H_r^{in}(z)&=&zG_{r}(H_r^{in}(z),1,H_u^{in}(z)),\\
  H_u^{in}(z)&=&zG_u(H_r^{in}(z),1,H_u^{in}(z)),\\
  H^{in}(z)&=&zG_0(H_r^{in}(z),1,H_u^{in}(z)).\label{eq.Ap.HHin}
\end{eqnarray}
Ref.~\cite{kenah2007second} showed that $\Pi=1-H^{out}(1) \leq
R=1-H^{in}(1)$, and hence the SIR does not map with link percolation
because this percolation process implies that $\Pi=R$. However, it was
shown in Ref.~\cite{kenah2007second} that for the case in which the
recovery time is constant, forward and backward branching are
equivalent and consequently $R = \Pi$.

\subsection{Non-homogeneous recovery time with homogeneous transmissibility}\label{ap.Nondeg}

In Ref.~\cite{kenah2007second}, the authors presented the main ideas
and equations to solve the SIR model with any recovery time $\tau_R$
distribution and an infection time $\tau_I$ that follows an exponential
distribution. They showed that for any recovery time distribution,
$\Pi \leq R$, and the equality holds when $\tau_R$ is
constant. Here we develop a toy-model in which the equality is valid
for heterogeneous recovery time $\tau_R$ distribution but with constant
transmissibility $T_{\tau_R}$. This case is relevant in our study because we
obtain that for a network with communities, there is not a strong
dependence between the transmissibility $T_{\tau_R}$ and $\tau_R$ (see
Fig.~\ref{fig.hist}c).

To study the effect on the final state of a heterogeneous $\tau_R$ distribution
with $T_{\tau_R}$ constant, we propose the following recovery time
distribution
\begin{eqnarray}\label{eq.aptrtoy}
  P(\tau_R)&=&0.5\delta_{\tau_R,2}+0.5\delta_{\tau_R,10} 
\end{eqnarray}
where $\delta$ is the Kronecker delta, and the infection time distribution
$P(\tau_I|\tau_R)$ is given by Table~\ref{eq.ap.Simap} where $\sigma \in [0,1]$.
\begin{table}[H]
\centering
   \begin{tabular}{|c|c|c|c|c|}
        \hline
         & $\tau_I=\infty$   & $\tau_I=1$   &  $\tau_I=2$   & $\tau_I=10$         \\ \hline
         $\tau_R=2$  & 1-$\sigma$   & 0.5$\sigma$   &  0.5$\sigma$  & 0                      \\ \hline
         $\tau_R=10$  & 1-$\sigma$   & 0.5$\sigma$   & 0 & 0.5$\sigma$                        \\ \hline
   \end{tabular}
\caption{Distribution $P(\tau_I|\tau_R)$ for $T_{\tau_R}$ constant}\label{eq.ap.Simap}
\end{table}
For this case, $T=T_{\tau_R}=\sigma$. For the purpose of comparison,
we also study a similar distribution $P(\tau_I|\tau_R)$ in which
$T_{\tau_R}$ is not constant (see Table~\ref{eq.ap.Nomap}).

\begin{table}[H]
\centering
   \begin{tabular}{|c|c|c|c|c|}
        \hline
         & $\tau_I=\infty$   & $\tau_I=1$   &  $\tau_I=2$   & $\tau_I=10$         \\ \hline
         $\tau_R=2$  & 1-$\sigma^{10}$   & 0.5$\sigma^{10}$   &  0.5$\sigma^{10}$  & 0                      \\ \hline
         $\tau_R=10$  & 1-$\sigma$   & 0.5$\sigma$   & 0 & 0.5$\sigma$                        \\ \hline
   \end{tabular}
\caption{Distribution $P(\tau_I|\tau_R)$ for non-homogeneous $T_{\tau_R}$}\label{eq.ap.Nomap}
\end{table}
Using the same recovery time distribution $P(\tau_R)$ as in the
previous case, the transmissibilities are: $T_{\tau_R=2}=\sigma^{10}$,
$T_{\tau_R=10}=\sigma$, and $T=0.5\sigma+0.5\sigma^{10}$.

\begin{figure}[H]
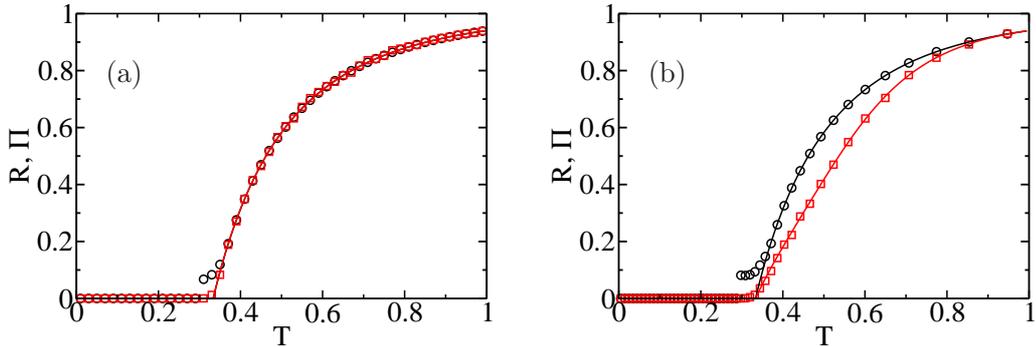

\vspace{0.5cm}
\begin{center}
\begin{overpic}[scale=0.25]{Fig13a.eps}
  \put(20,55){(a)}
\end{overpic}
\hspace{0.5cm}
\begin{overpic}[scale=0.25]{Fig13b.eps}
  \put(20,55){(b)}
\end{overpic}
\end{center}
\caption{Mapping between link percolation and the SIR model with
  heterogeneous recovery distribution for ER networks with $\langle k
  \rangle=3$. The panels show $R$ (black) and $\Pi$ (red) as a
  function of the total transmissibility $T$ for a heterogeneous
  recovery distribution given by Table~(\ref{eq.ap.Simap}) (panel a) and
  Table~(\ref{eq.ap.Nomap}) (panel b). The lines correspond to the
  theoretical solutions of
  Eqs.~(\ref{eq.Ap.kkin})-(\ref{eq.Ap.HHin}), and the symbols to
  simulations. The simulations were performed over $10^4$ network
  realizations with $N=10^4$ and $s_c=600$. The disagreement between the
  theoretical curves and the simulations around the critical point is
  due to finite size effects and the value of $s_c$ which cannot
  distinguish an epidemic from an outbreak near $T=T_c$ (see
  Appendix~\ref{ap.sc}).}\label{fig.ApMapSiNo}
\end{figure}

In Fig.~\ref{fig.ApMapSiNo}a-b, we show $R$ and $\Pi$
obtained from the theory (Eqs.~(\ref{ap.eq.Gxyu})-(\ref{eq.Ap.HHin}))
and simulations for the recovery and infection time distributions in
Eqs.~(\ref{eq.aptrtoy}) and Table~(\ref{eq.ap.Nomap}). Our results confirm that
for a constant $T_{\tau_R}$, the probability of an epidemic is equal
to the fraction of recovered nodes (Fig.~\ref{fig.ApMapSiNo}a) even if
$P(\tau_R)$ is heterogeneous, while for non-constant $T_{\tau_R}$,
$\Pi<R$ (Fig.~\ref{fig.ApMapSiNo}b). Thus, it
is expected that for a weak dependency between $T_{\tau_R}$ and
$\tau_R$, link percolation is a good approximation of the SIR
model, as shown in Fig.~\ref{fig.probpand}.

\bibliography{bib}

\end{document}